\documentclass{iau-JDSS}
\usepackage{graphicx}

\begin{document}

\title[Rotation of classical bulges] 
{Rotation of classical bulges during secular evolution of barred galaxies}

\author[Kanak Saha]   
{Kanak Saha \& Ortwin Gerhard%
}

\affiliation{Max-Planck-Institut f\"ur Extraterrestrische Physik, Giessenbachstra\ss e, D-85748 Garching, Germany, \break email: saha@mpe.mpg.de}

\pubyear{2012}
\volume{Volume 15}  
\pagerange{119--126}
\date{?? and in revised form ??}
\setcounter{page}{119}
\jname{Highlights of Astronomy, Volume 15}
\editors{Thierry Montmerle, ed.}

\maketitle

\begin{abstract}
Bar driven secular evolution plays a key role in changing the
morphology and kinematics of disk galaxies, leading to the formation
of rapidly rotating boxy/peanut bulges. If these disk galaxies also 
hosted a preexisting classical bulge, how would the secular evolution
influence the classical bulge, and also the observational properties.

We first study the co-evolution of a bar and a preexisting
non-rotating low-mass classical bulge such as might be present in
galaxies like the Milky Way. It is shown with N-body simulations that
during the secular evolution, such a bulge can gain significant
angular momentum emitted by the bar through resonant and stochastic
orbits. Thereby it transforms into a cylindrically rotating,
anisotropic and triaxial object, embedded in the fast rotating boxy
bulge that forms via disk instability (Saha et al. 2012). 
The composite boxy/peanut bulge also rotates cylindrically.

We then show that the growth of the bar depends only slightly on the rotation
properties of the preexisting classical bulge. For the initially
rotating small classical bulge, cylindrical rotation in the resulting
composite boxy/peanut bulge extends to lower heights (Saha \& Gerhard 2012). 
More massive classical bulges also gain angular 
momentum emitted by the bar, inducing surprisingly large rotational 
support within about 4~Gyrs (Saha et al. in prep) .

\end{abstract}

\begin{figure}[h!]
\centering
\includegraphics[width=14cm,angle=0]{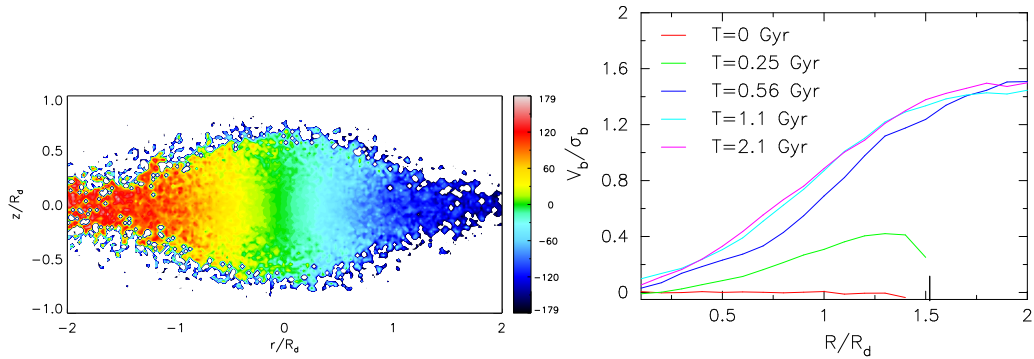}
  \caption{Line-of-sight velocity map showing cylindrical rotation of the
classical bulge particles at time $2.1$~Gyr in the Saha et al. (2012) model (left). 
Variation of rotation velocity with radius and time during secular evolution (right).}
\end{figure}

\end{document}